\documentclass[twocolumn]{aastex63}

\usepackage{amsmath}
\usepackage{array}
\usepackage{url}
\usepackage{colortbl}
\newcommand{\DMunit}{$\rm pc\,cm^{-3}$}

\newcommand{\toa}{TOA}
\def\PSR{J2241$-$5236}

\received{December 2, 2021}
\revised{March 22, 2022}
\accepted{April 7, 2022}
\submitjournal{ApJ Letters}

\shorttitle{Frequency-dependent DM towards {\PSR}}
\shortauthors{Kaur et al.}
\begin{document}

\title{Detection of frequency-dependent dispersion measure toward the millisecond pulsar {\PSR} from contemporaneous wide-band observations}

\correspondingauthor{N. D. R. Bhat}
\email{ramesh.bhat@curtin.edu.au}

\author[0000-0003-4879-1019]{Dilpreet Kaur}
\affiliation{International Centre for Radio Astronomy Research (ICRAR), Curtin University, Bentley, WA 6102, Australia}
\affiliation{CSIRO Astronomy and Space Science, PO Box 76, Epping, NSW 1710, Australia}

\author[0000-0002-8383-5059]{N. D. Ramesh Bhat}
\affiliation{International Centre for Radio Astronomy Research (ICRAR), Curtin University, Bentley, WA 6102, Australia}

\author[0000-0002-9618-2499]{Shi Dai}
\affiliation{CSIRO Astronomy and Space Science, PO Box 76, Epping, NSW 1710, Australia}
\affiliation{Western Sydney University, Locked Bag 1797, Penrith South DC, NSW 1797, Australia}

\author[0000-0001-6114-7469]{Samuel J. McSweeney}
\affiliation{International Centre for Radio Astronomy Research (ICRAR), Curtin University, Bentley, WA 6102, Australia}

\author[0000-0002-7285-6348]{Ryan M. Shannon}
\affiliation{Centre for Astrophysics and Supercomputing, Swinburne University of Technology, P.O. Box 218, Hawthorn, VIC 3122, Australia}
\affiliation{ARC Centre of Excellence for Gravitational Wave Discovery (OzGrav), Swinburne University of Technology, Hawthorn, Australia}

\author[0000-0002-6631-1077]{Sanjay Kudale}
\affiliation{National Centre for Radio Astrophysics, Tata Institute of Fundamental Research, Pune 411 007, India}

\author[0000-0003-2519-7375]{Willem van Straten}
\affiliation{Institute for Radio Astronomy \& Space Research, Auckland University of Technology, Private Bag 92006, Auckland 1142, New Zealand}




%revised text RB 
\begin{abstract}
Making precise measurements of pulsar dispersion measures (DMs) and applying suitable corrections for them is amongst the major challenges in high-precision timing programmes such as pulsar timing arrays (PTAs). While the advent of wide-band pulsar instrumentation can enable more precise DM measurements and thence improved timing precision, it also necessitates doing careful assessments of frequency-dependent (chromatic) DMs that was theorised by \cite{Cordes2016}.
Here we report the detection of such an effect in broadband observations of the millisecond pulsar PSR {\PSR}, a high-priority target for current and future PTAs. 
The observations were made contemporaneously  using the wide-band receivers and capabilities now available at the Murchison Widefield Array (MWA), the upgraded Giant Metrewave Radio Telescope (uGMRT) and the Parkes telescopes, thus providing an unprecedentedly large frequency coverage from 80\,MHz to 4\,GHz. Our analysis shows the measurable changes in DM that scale with the observing frequency ($\nu$) as $\rm \delta DM \sim \nu^{2.5 \pm 0.1}$. We discuss the potential implications of such a frequency dependence in the measured DMs, and the likely impact on the timing noise budget, and comment on the usefulness of low-frequency observations in advancing PTA efforts.
\end{abstract}

\keywords{pulsars: general --- pulsars: individual (PSR {\PSR}) --- instrumentation: interferometers ---ISM: general}


\section{Introduction} 
\label{sec:intro}
%Pulsar timing array (PTA) experiments require sub-microsecond accuracies in the measured arrival times for the eventual detection of low-frequency (nanohertz) gravitational waves \citep[GWs;
%Revised RB
Pulsar timing array (PTAs) rely on high-precision measurements of pulse arrival times for a large number of pulsars over long time spans for the eventual detection of low-frequency (nanohertz) gravitational waves \citep[GWs;
e.g.,][]{Foster&Backer1990, Manchester2013PPTA, Demorest2013, Bailes2018, VanHaasteren2011, Janssen2015}.
To reach the sensitivity required to detect nanohertz GWs \citep{Cordes&Shannon2010, Lentati2013, Levin2016, Lam2016, Shannon2014, Foster2015}, PTAs around the world strive to achieve timing precisions of $\sim$100-200\,ns in pulse times of arrival (TOAs). This goal motivates careful assessment of various astrophysical and instrumental effects that contribute to the timing noise budget, and the development of strategies to mitigate them in the timing data.

The advent of wide-band receivers, such as the ultra-wideband low-frequency receiver \citep[UWL; 704-4032\,MHz;][]{Hobbs2020} at the 64-m Parkes telescope (also known as \emph{Murriyang}), the L-band receiver at the MeerKAT (580-1670\,MHz) telescope \citep{Bailes2020}, and the upgraded Giant Metrewave Radio Telescope \citep[uGMRT; 120-1450\,MHz;][]{Gupta2017}, is highly promising for improved timing precision in PTA observations. However, these large bandwidths also necessitate additional considerations; for example, correcting for the interstellar medium effects on pulsar signals such as dispersion, scintillation and multi-path scattering, 
%all of which scale strongly with the inverse of the observing frequency.
%RB revised
all of which scale strongly with the observing frequency.

The dispersive delay is proportional to the integrated column density of free electrons along the line of sight of the pulsar. The time delay for a pulse observed at radio frequency $\nu$, compared to a (hypothetical) pulse at infinite frequency can be approximated as $t_{\rm DM} = \mathcal{D}\,{\rm DM}/\nu^{2}$, where $\mathcal{D} = e^2/(2\pi m_e c)$ is the dispersion constant, $e$ and $m_e$ are the charge and mass of the electron, $c$ is the speed of light, and DM is the dispersion measure, 
%or 
%SJM suggested text
traditionally defined as the 
integral of the electron density ($n_e$) along the line of sight (LOS). 
%For PTAs, it is a time-varying quantity, largely due to the pulsar's proper motion (typically $\rm \sim 50$-$\rm 100\,km\,s^{-1}$), 
%Revised RB
For PTAs, it is a time-varying quantity, primarily due to the pulsar's large space velocities (typically $\rm \sim 50$-$\rm 100\,km\,s^{-1}$; \citealp{Hobbs2005pm}), 
as a result of which different parts of the interstellar medium (ISM) are sampled by the observations. Since spatial variations in electron density are present on a wide range of scales ($\sim10^{6}$ to $10^{13}$\,m or even larger; \citealp{Lam2015, Cordes1998}), PTA observations require measuring and correcting for DM at every observing epoch. Measuring DMs accurately and applying suitable corrections for their temporal variability has been an integral part of pulsar timing array experiments \citep{Jones2017, Lam2016, Keith2013, Lee2014, You2007}.

Aside from the time variability in DM, there may also be a frequency dependence; sometimes referred to as `chromaticity' in DM. 
The importance of such a subtle effect was first acknowledged by \citet{Ramach2006} in their analysis of DM variations of the first-discovered millisecond pulsar (MSP) B1937$+$21. 
More recently, \citet{Cordes2016} provided a detailed theoretical account of the underlying physics, and discussed possible implications in the wider context of PTAs. 
The effect arises as a result of multi-path scattering by the electron density fluctuations present in the ISM. Since the radiation received at a given frequency depends on the size of the scattering disk, which scales as the square of the observing wavelength ($\rm \theta_{s} \propto \lambda^{2}$), the volume of the ISM (or path lengths) sampled at lower frequencies can be significantly larger than those at higher frequencies, especially when observations are made over very large bandwidths (hundreds of MHz to a few GHz). 
%\citet{Cordes2016} formulated this effect theoretically, and demonstrated through simulations. 
Such a frequency dependence in DM can give rise to subtle differences in DMs as measured in observations, depending on the observing frequency and the bandwidths over which such measurements are made. 
%As an example, a DM variation of $\sim 10^{-5}$\,{\DMunit} (e.g., a pulsar at a distance of $\sim$1 kpc and DM $\lesssim$30\,{\DMunit}) can give rise to $\sim$110\,ns of timing noise at standard timing frequency bands of $\sim$\,1-2\,GHz. This  can be of the order of few microseconds for high DM pulsars and over wider observing bandwidths \citep{Cordes2016}. 
%RB revised in response to the referee comment - for improved clarity 
For example, as per the formalism presented in
\cite{Cordes2016}, we may expect an rms DM variation $\sim 10^{-5}$\,{\DMunit} for a pulsar at a distance of $\sim$1\,kpc and DM $\lesssim$30\,{\DMunit}, which may result in a few hundred nanoseconds of timing noise at the standard timing frequency bands ($\sim$\,1-2\,GHz). This can be of the order of up to a few microseconds for higher DM pulsars and over wider observing bandwidths \citep{Cordes2016}. The investigation of such subtle effects are particularly important in the era of wideband precision pulsar timing, when the observations are routinely made over large bandwidths of the order of hundreds of MHz to a few GHz.

To date, there have been limited observational investigations to study this effect. 
% RB: additonal text following the comment from Shi who pointed out the 1937 DM paper
In their analysis of 20 yr timing data on the MSP B1937$+$21, \citet{Ramach2006} interpreted parts of their analysis (i.e. observations at 327 and 610 MHz) in terms of frequency dependent variations in DM. 
More recently, 
\cite{Donner2019} studied observations taken with three German LOng-Wavelength (GLOW) stations, which are part of the LOw-Frequency ARray (LOFAR). They presented 3.5\,years of weekly observations of PSR J2219$+$4754, a long-period pulsar with a DM of 43.5\,{\DMunit}. The frequency-dependent DM trends seen towards this pulsar (at 100-200\,MHz) were attributed to extreme scattering events (ESEs) caused by a discrete cloud of size $\sim$20\,AU and $n_e$ $\sim10$ $\rm cm^{-3}$. 
These results were revisited by \citet{Lam2020}, who concluded that the DM variations are due to the turbulent ISM and no ESE occurred along the LOS. 
They also comment on the suitability of long-period pulsars for investigating frequency-dependent DM, given their distinct emission characteristics compared to those of millisecond pulsars that are used for timing-array experiments.

In this paper, we present the measurements of frequency-dependent DMs toward the millisecond pulsar PSR {\PSR}. 
%Its low DM (11.41151\,{\DMunit}) and short pulse period (2.18\,ms) make it an important target for PTAs. 
%Revised RB
Its low DM (11.41151\,{\DMunit}) and short pulse period (2.18\,ms), along with its brightness and low levels of timing noise make it an important target for PTAs \citep{Keith2011,Dai2015,Kaur2019,Parthasarathy2021}.
%The pulsar is in a 3.5\,hour, almost-circular orbit with a low-mass ($\sim$\,0.012\,$\rm M_{\odot}$), black-widow type companion \citep{Keith2011}. 
%Revised RB
The pulsar is in a 3.5\,hour, almost-circular orbit with a low-mass ($\sim$\,0.012\,$\rm M_{\odot}$), black-widow type companion, but with no eclipses seen \citep{Keith2011}. 
For this work, observations were carried out \emph{contemporaneously} using the Murchison Widefield Array (MWA; \citealp{Tingay2013,Wayth2018}), the upgraded Giant Metrewave Radio Telescope (uGMRT), and the new ultra-wide band low-frequency receiver (UWL) at the Parkes radio telescope, thus providing frequency coverage from 80\,MHz to 4\,GHz.
The remainder of the paper is organised as follows.
In $\mathsection\,2$ we describe the details of our observations. In $\mathsection\,3$ and $\mathsection\,4$ we summarize data processing and analysis, and in $\mathsection\,5$ we present our results. Our conclusions are summarized in $\mathsection\,6$.

\section{Observations} \label{sec:obs}
Observations of PSR {\PSR} were made contemporaneously (i.e. within a $\sim$24-hour duration) using the MWA, uGMRT and Parkes.
This pulsar is ideal for our analysis since it has a narrow pulse profile and 
very little profile evolution across the observed frequency range,
as discussed below. 
The pulsar was observed at multiple (three) epochs with all three telescopes. At two of the epochs, it was observed within a 24\,hour block using all three telescopes, while at one of the epochs, only two telescopes (uGMRT and Parkes) were available. Observational details are summarised in Table~\ref{tab:obs_summary}, and are further elaborated below.

\begin{deluxetable*}{cccccccc}
\tablenum{1}
\tablecaption{Summary of observational parameters\label{tab:obs_summary}}
\tablewidth{1pt}
\tablehead{
%\colhead{Date of obs} & \multicolumn5c{Telescope/Observed band (MHz)} \\
%\cline{2-7}
\colhead{MJD} & \colhead{Telescope/receiver} & Frequency range & \colhead{$\rm \Delta t$} & \colhead{$\rm T_{obs}$} & \colhead{$\rm N_{scan}$} &
\colhead{$\rm N_{chan}$} & \colhead{$f_{\rm template}$} \\
\colhead{} & \colhead{} & \colhead{(MHz)} & \colhead{($\mu$s)} & \colhead{(minutes)} & \colhead{} & \colhead{} & \colhead{(MHz)} 
}
\startdata
$58671^{\dagger}$ & MWA VCS & 80-220 & 0.78 & 80 & 8 & 24 & 150  \\
$58680^{\dagger}$ & MWA VCS & 80-220 & 0.78 & 40 & 4 & 24 & 150 \\
58794 & uGMRT Band 3 & 300-500 & 10.24&90  & 3 & 32 & 400 \\
58795 & Parkes UWL & 704-4032& 2.128&65  & 3 & 26 & 1300 \\
{}& MWA VCS & 80-220 &0.78 &45  & 1 & 24 & 150 \\
58796 & Parkes UWL & 704-4032 &2.128 &65  & 3 & 26 & 1300 \\
58799 & uGMRT Band 3 & 300-500 &10.24 &105  & 4 & 32 & 400 \\
58800 & Parkes UWL & 704-4032 & 2.128&65  & 3 & 26 & 1300 \\
58816 & Parkes UWL & 704-4032 & 2.128&65  & 3 & 26 & 1300 \\
{}& uGMRT Band 4 & 550-750 &10.24 &65  & 2 & 32 & 650 \\
{}& MWA VCS & 80-220 &0.78 &45 & 1 & 24 & 150 \\
58817 & Parkes UWL & 704-4032 & 2.128&65 & 2 & 26 & 1300 \\
\enddata
$^{\dagger}$\,Dedicated observations with the MWA to sample the full 3.5-hr orbit, across the 80-220 MHz range (see \S~2.1.1 for further details).
\tablecomments{$\rm \Delta t$ is the time resolution, $\rm T_{obs}$ is the total observing duration, $\rm N_{scan}$ is the number of independent observations (scans) each of $\rm T_{obs}/N_{scan}$ duration, $\rm N_{chan}$ is the number of channels (sub-bands) used to measure the TOAs, and $f_{\rm template}$ 
is radio frequency of the observations from which an analytic template was derived.}
\end{deluxetable*}

%%%%%
\begin{figure}
\includegraphics[width=3.39in,angle=0]{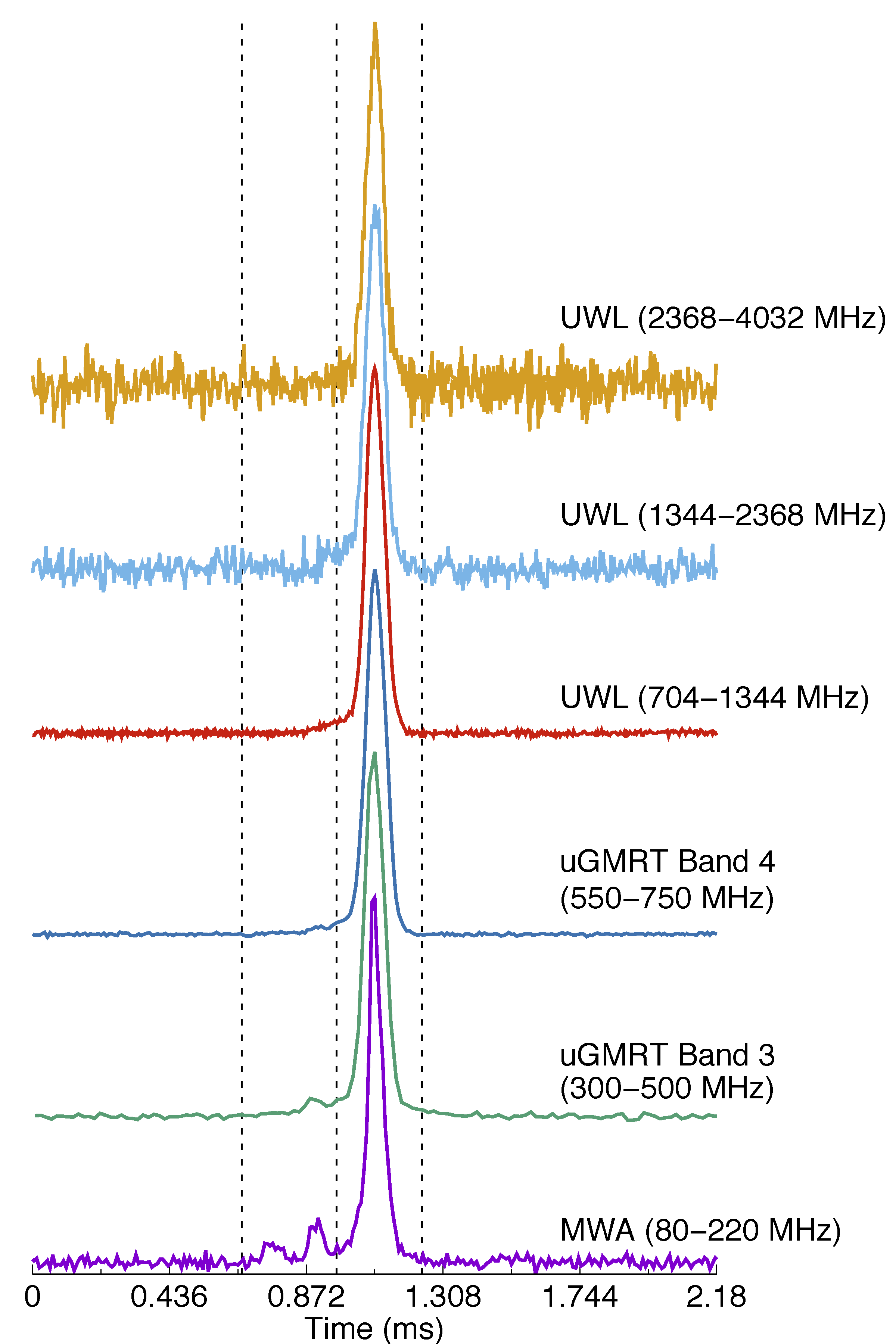}
\caption{Integrated pulse profiles of PSR {\PSR} at frequencies from 150\,MHz to 4032\,MHz. 
The Parkes profiles span the UWL band from 704 to 4032 MHz, which is sub-divided into three segments. Very little profile evolution is seen across the large frequency range spanned by our observations.
\label{fig:1}}
\end{figure}
%%%%%
%%\begin{figure}
%%\gridline{\fig{Fig1.png}{0.4\textwidth}{}}
%%\caption{Integrated pulse profiles of PSR {\PSR} at frequencies from 150\,MHz to 4032\,MHz.  The Parkes profiles span the UWL band from 704 to 4032 MHz, which is sub-divided into three segments. Very little profile evolution is seen across the large frequency range spanned by our observations.
%%\label{fig:1}}
%%\end{figure}

\subsection{The MWA}
Observations were made over a total of four epochs, two epochs to sample the full orbit (described in Section \ref{sec:mwa_orbital_coverage}) and another two contemporaneously (separated by three weeks) with other telescopes, using the MWA's voltage capture system \citep[VCS;][]{Tremblay2015}. The VCS records 24 (1.28\,MHz wide) coarse channels, each of which has been finely channelized to 10\,kHz. These data were recorded in both polarizations, from all 128 tiles. The 30.72\,MHz observing bandwidth of the MWA can be flexibly placed (e.g., $12\times2.56$\,MHz coarse channels) anywhere across the nominal operating frequency range (80-300\,MHz). For the observations presented in this paper, a frequency set-up similar to that described in \citet{Kaur2019} was employed, i.e., a distributed $24\times1.28$\,MHz channel mode, covering a frequency range from 80\,MHz to 220\,MHz.

\subsubsection{Orbital coverage}
\label{sec:mwa_orbital_coverage}
Since the pulsar is in a binary system, it was important to examine the data for any orbital DM variations. The ability to sample a full orbit of the pulsar in a single day can help to minimise any temporal variations in DM caused by the ISM. With the VCS, the maximum recording time is however limited to $\sim90$\,minutes, which results in a data volume of $\sim 50$\,TB (data rate = $\rm 7.78\,GB\,s^{-1}$). The data were recorded in multiple 10\,minute recordings at $\sim$15\,minute intervals between the recordings. This allowed us to optimally use the available 90 minutes of recording time to sample the full binary orbital period of 3.5 hours. A ``picket-fence" mode of observation was adopted in order to achieve a large frequency coverage, but with a modified frequency-channel selection where the 30.72\,MHz bandwidth was divided into five uneven sub-bands of 2$\times$7.68\,MHz and 3$\times$5.12\,MHz centred at 83.84\,MHz, 127.36\,MHz, 155.52\,MHz, 187.52\,MHz, and 218.24\,MHz respectively. This strategy was chosen to increase the sub-bandwidth in order to compensate for the reduced signal-to-noise ratio (S/N) associated with breaking up the recording time into 10\,minute observations. The main purpose was to ensure a sufficiently high S/N to detect the pulsar in each of the sub-bands in 10 minute recordings (as compared to the typical observing time of $\sim$\,1\,hour for this pulsar that was used for the work presented in \citealp{Kaur2019}). In summary, by making optimal use of available resources in bandwidth and time, the full 3.5 hour orbit was sampled in a total of eight 10\,minute recording sessions.

\subsection{The uGMRT}
We conducted observations using the upgraded GMRT \citep[uGMRT;][]{Gupta2017}. Two observations were made in Band 3, which covers from 300 to 500\,MHz, and one observation in Band 4, which covers 550 to 750\,MHz. Observations were made using the phased array total intensity mode.
Depending upon the observing epoch, the number of antennas used varied from 24 to 28. The data were then coherently combined to generate spectral (channelised) voltages time series data. 
These data were processed in real-time using the coherent dedispersion processing pipeline
of the uGMRT wideband backend \citep[GWB;][]{Reddy2017}. For all observations, data were recorded in 512$\times$0.390\,MHz coherently-dedispersed filter-bank format with $10.24\,\mu$s time resolution. 
In order to avoid de-phasing due to ionospheric effects or antenna gain changes, we performed re-phasing of the array periodically, at an average interval of $\sim30$\,minutes, using a nearby phase calibrator 2225$-$049, which has a flux density of 15\,Jy at P-band (400\,MHz). The incoming data were then processed by the GWB, and data were recorded with 256 phase bins across the pulse period, for both the uGMRT Band 3 and Band 4 observations.

\subsubsection{Orbital coverage}
Given the far southern declination of the pulsar, and the geographic location of the uGMRT, PSR {\PSR} is visible for a maximum of 2.1\,hours on any particular day. This covers approximately 60\% of the orbit.
In order to span the full 3.5\,hour orbit, two successive observations were carried out, five\,days apart. The first observation on MJD 58794 covered 
the orbital phase range from $\sim 0.6 P_b$ to 0.8\,$P_b$, where $P_b$ is the orbital period, 
while the second observation on MJD 58799 
covered from $\sim$\,0.8 to 1.0 and 0.0 to $\sim$\,0.3 phase ranges. The two observations thus partially covered the orbit, with no coverage for $\sim$\,0.3\,$P_b$ to $\sim$\,0.6\,$P_b$ in the orbital phase range. Data recording was interrupted for a short duration ($\sim$\,5 minutes) to accommodate the phase calibrators and re-phasing of the array.

\subsection{Parkes}
Pulsar {\PSR} is regularly monitored at Parkes as part of the ongoing Parkes pulsar timing array \citep[PPTA;][]{Manchester2013PPTA} project. The use of the ultra-wideband low-frequency receiver (UWL; \citealp{Hobbs2020}) provides simultaneous frequency coverage from 704 to 4032\,MHz. It records signals in three adjacent radio frequency bands; 704-1344\,MHz, 1344-2368\,MHz and 2368-4032\,MHz. The entire band is sampled by digitisers for both polarisations. Digitised data from the telescope's focus cabin are transported to the signal pre-processing units where the entire band is channelised into 26 contiguous sub-bands, each 128\,MHz wide, and for each polarisation.

Parkes observations used the pulsar fold mode where data are synchronously folded modulo the pulse period, with 1024 phase bins in each of the 1\,MHz channels (and hence a time resolution $\approx$2.128\,$\mu$s), and integrated for 30\,s before written out to disk. Data were coherently dedispersed at a DM of 11.41151\,{\DMunit}, and the Stokes parameters were recorded. 

\begin{figure*}
\gridline{\fig{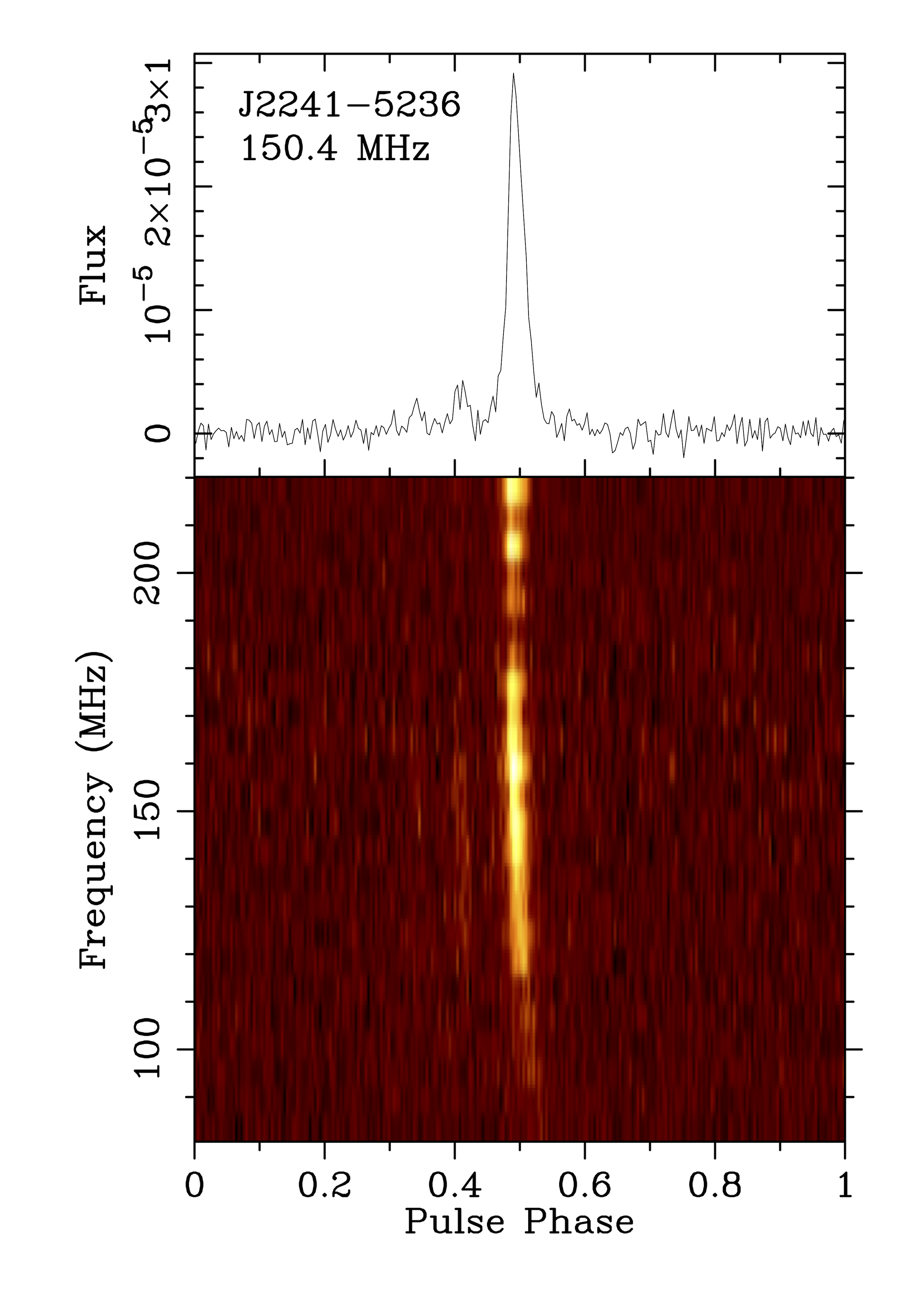}{0.24\textwidth}{(a)}
          \fig{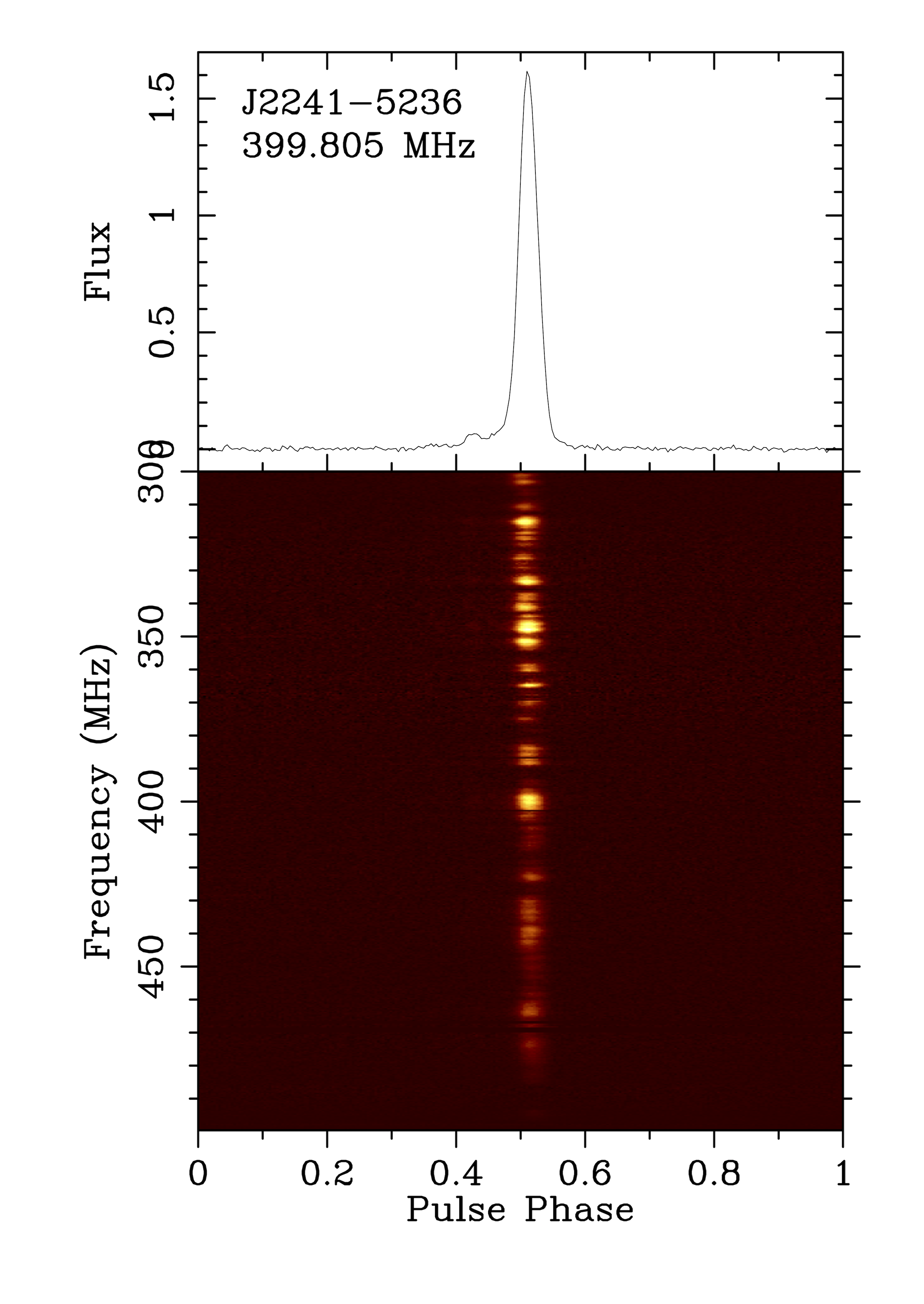}{0.24\textwidth}{(b)}
          \fig{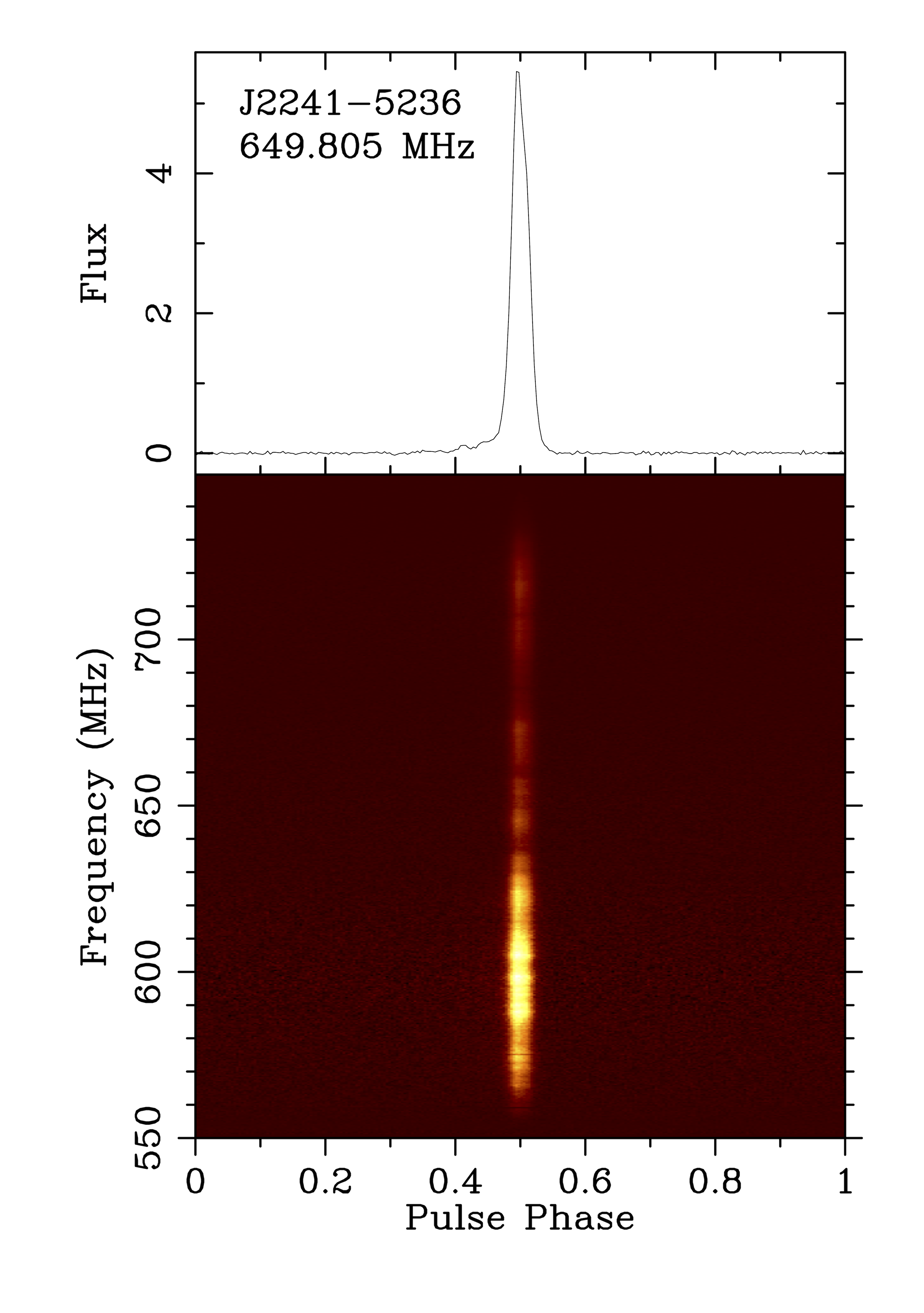}{0.24\textwidth}{(c)}
          \fig{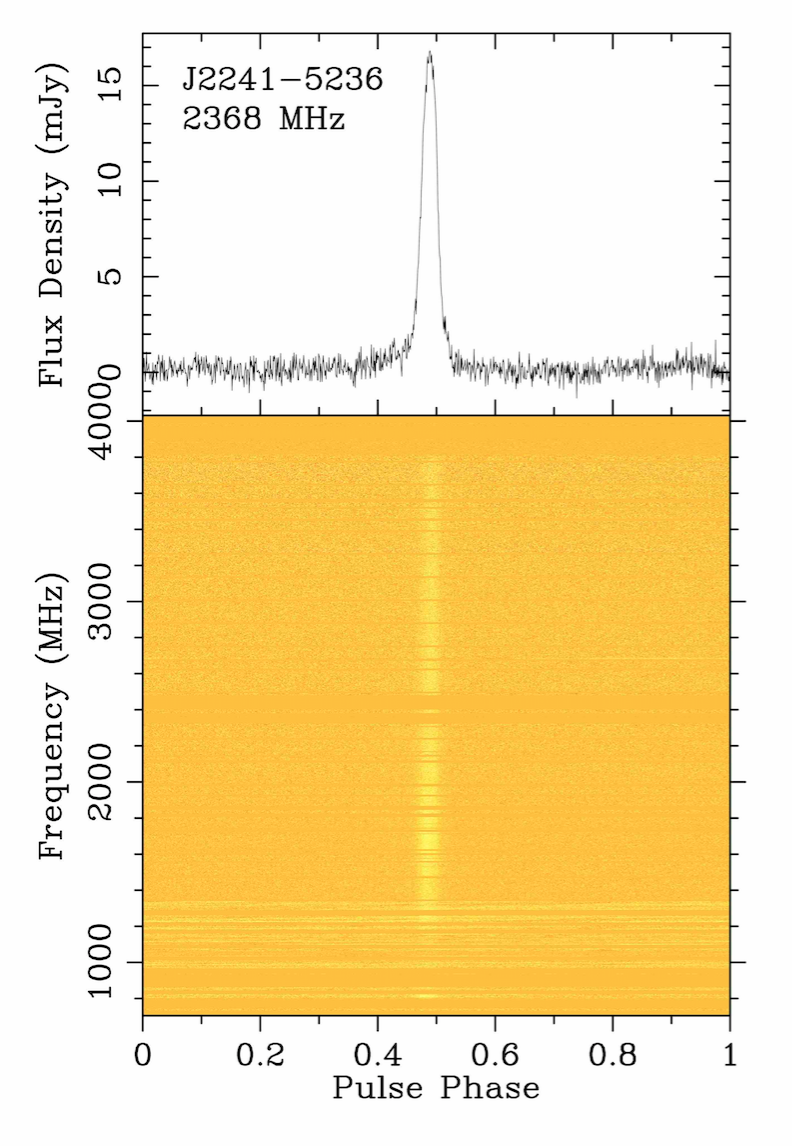}{0.24\textwidth}{(d)}
          }
\caption{Coherently de-dispersed pulsar detections shown as average profiles integrated over frequency (top panels) and frequency vs. pulse phase waterfall plots (bottom panels): (a) detection across the MWA's  80-220\,MHz, band (centred at 150.4\,MHz), with a time resolution of $\approx 4.25\,\mu$s; (b) detection across the uGMRT Band 3 (300-500\,MHz, centred at 399.805\,MHz), with $\approx 10.24\,\mu$s time resolution, (c) detection across the uGMRT Band 4 (550-750\,MHz, centred at 649.805\,MHz), with a time resolution of $\approx 10.24\,\mu$s, and (d) detection across the entire UWL band (704-4032\,MHz, centred at 2368\,MHz) with $\sim 2.128\,\mu$s time resolution. Only UWL data are calibrated to an absolute flux density scale (and hence in mJy), whereas the flux scale is in arbitrary units for MWA and uGMRT data.  
\label{fig:style_plots}}
\end{figure*}

\definecolor{Gray}{gray}{0.9}
\begin{deluxetable*}{ccccccc}
\tablenum{2}
\tablecaption{Summary of DM measurements (in {\DMunit})\label{tab:MWA_observations}}
\tablewidth{1pt}
\tablehead{
%\colhead{Date of obs} & \multicolumn5c{Telescope/Observed Band} \\
%\cline{2-7}
\colhead{MJD} & \colhead{MWA} & \colhead{uGMRT} & \colhead{uGMRT} &
\colhead{Parkes UWL} & \colhead{Parkes UWL} & \colhead{Parkes UWL} \\
\colhead{} & \colhead{VCS} & \colhead{Band 3} & \colhead{Band 4} & \colhead{low band} & \colhead{mid band} & \colhead{high band} \\
\colhead{} & \colhead{(80-220 MHz)} & \colhead{(300-500\,MHz)} & \colhead{(550-750\,MHz)} & \colhead{(704-1344\,MHz)} & \colhead{(1344-2368\,MHz)} & \colhead{(2368-4032\,MHz)}
}
\startdata
58671& 11.411355(2) & & & & & \\
58680& 11.411371(6) & & & & & \\
58794&  & 11.41134(2) & & & & \\
\rowcolor{Gray}
     &  & 11.412037(6)&&&& \\
58795& 11.411327(4) &{}&{}& 11.410(2) & 11.4075(8) & 11.397(4) \\
\rowcolor{Gray}
& 11.412142(6)&&&11.4107(3)&11.4089(5)&11.398(3) \\
58796&  &   &   & 11.4099(4) & 11.411(1) & 11.397(4) \\
\rowcolor{Gray}
&  &   &   & 11.4109(7)&11.409(1)&11.399(2) \\
58799&  & 11.41131(2) &   &  &  & \\
\rowcolor{Gray}
&  & 11.412301(2)& &  &  & \\
58800&  &  &  & 11.41119(2) & 11.4084(8) & 11.397(3) \\
\rowcolor{Gray}
&  &  &  & 11.41131(6)&11.4092(4)&11.407(3) \\
58816& 11.411301(3) &  & 11.41112(2) & 11.4111(4) & 11.408(1) & 11.393(3) \\
\rowcolor{Gray}
&11.411340(4)&  & 11.41115(1)&11.410(1)&11.409(1)&11.394(4) \\
58817&  &  &  & 11.41132(9) & 11.4086(9) & 11.389(5) \\
\rowcolor{Gray}
&  &  &  &11.41136(9)&11.4091(8)&11.396(4) \\
\enddata
\tablecomments{Observational details and measured DM values from contemporaneous broadband data used for our analysis. The DM values have been measured using two different techniques, i.e. using the analytic template (narrow-band timing) and wideband timing (the latter is shown in shaded grey). The frequency range covered from each telescope is indicated in the top parentheses. The uncertainties in DMs are given in parentheses and correspond to the least significant digit. For MWA data, the DM precision is of the order of $10^{-6}$\,{\DMunit}, whereas it is $\sim \, 10^{-5}$\,{\DMunit} for the uGMRT bands (Bands 3 and 4) and is in the range $\sim$ $10^{-3}$ to $10^{-5}$\,{\DMunit} for the Parkes UWL band.}
\end{deluxetable*}

\section{Data processing and analysis}
\subsection{The MWA}
The MWA makes use of the VCS and the post-processing chain \citep{Bhat2016, Mcsweeney2017}. Data calibration and processing were performed on the Galaxy cluster at the Pawsey Supercomputing Centre. 
The signals from each tile can be either incoherently summed, or combined coherently, to generate a phased-array beam.   This process incorporates a model of the polarimetric response and complex gains of each of the tiles, including both cable and geometric delays \citep{Ord2019}. 
The phase model for the array was determined (the data are calibrated using this model as described earlier)
using one of the standard calibrators (e.g., 3C444), recorded in pointed observations prior to pulsar observations. Amplitude and phase calibration solutions were generated for each frequency sub-band using the real time system \citep[RTS;][]{Mitchell2008}. A coherent tied-array beam is produced by phasing up of the signals from all tiles with good calibration solutions,
essentially following the procedures as described in \citet{Bhat2016} and \citet{Ord2019}.
The high time resolution ($\approx1\,\mu$s) data were recovered using an enhanced capability of the beamformer software, which performs an inversion of the polyphase filterbank operation, the implementation of which is detailed in \citet{Mcsweeney2020}. Coherently de-dispersed data were then generated using the {\tt DSPSR} pulsar package \citep{van2011DSPSR} 
and average profiles were written in the PSRFITS (or Timer) file format
with $0.78\,\mu$s time resolution (see also \citealp{Kaur2019}). All the processing, including calibration and beamforming, was carried out at the Pawsey supercomputing facility.

On MJD 58671 and 58680, standard calibrator data were not available; therefore, for observations made on these dates, we adopted a somewhat non-standard approach.  Calibration solutions were constructed by selectively choosing parts of multiple different calibration observations that were recorded using frequency configurations that are not identical to the configuration of our pulsar observations.
%Even though the related procedure was tedious, the data were successfully calibrated, thereby demonstrating the efficacy of our approach. 
%Revised RB
Even though the related procedure was tedious, the data were successfully calibrated, resulting in pulsar detections with signal-to-noise ratios that are comparable to those obtained with a standard calibration process, thereby demonstrating the efficacy of our approach. 

\subsection{The uGMRT}
The uGMRT is located in a radio-frequency interference (RFI) hostile environment, because of which data are typically RFI contaminated. Although online RFI mitigation is possible with the uGMRT system, that was not used for our observations. All RFI excision analysis was performed offline using the {\tt pazi} routine of {\tt PSRCHIVE} by identifying and flagging the problematic channels. uGMRT Band 3 data are typically more prone to RFI, resulting in $\sim 20\%$ excised data, whereas only $\sim 10\%$ data were excised in Band 4.

\subsection{Parkes}
Parkes data reduction closely followed the procedures described in \cite{Hobbs2020} and \cite{Dai2019}. In short, we removed 5\,MHz of the bandpass at each edge of the 26 sub-bands to mitigate aliasing. The ``Coastguard" package was used 
to automatically excise RFI
by examining data in both time and frequency. Pulsed noise signal recorded before each observation, together with observations of the radio galaxy 3C 218 and PSR J0437$-$4715, were used to calibrate the differential and absolute gains and polarimetric response of the receiver \citep[cf.][]{Van2004}. After the calibration, data were further RFI cleaned using the {\tt pazi} routine of {\tt PSRCHIVE} by carefully (manually) flagging the problematic channels. The calibrated and RFI cleaned data were then integrated in time
for further timing and DM analysis.

\begin{figure}
\gridline{\fig{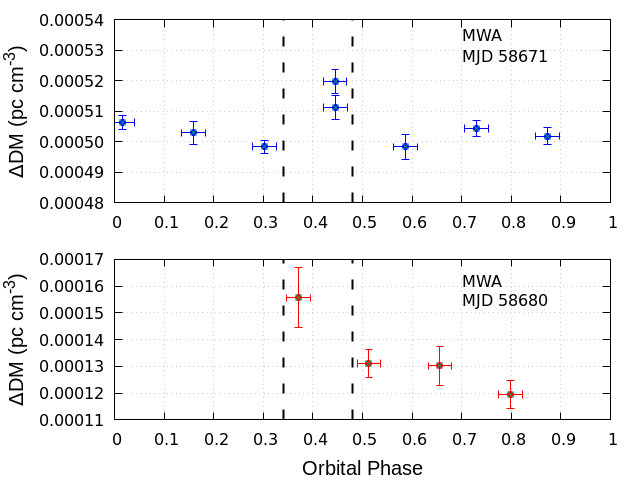}{0.50\textwidth}{}}
%\gridline{\fig{Fig3.png}{0.45\textwidth}{}}
\vspace*{-0.5cm}
\caption{DM variation vs. orbital phase, from dedicated MWA observations. The observing date (MJD) is indicated in the panel 
and, on each day, the measurements were made within a single (3.5-hr) orbital period. 
Top panel shows the DM measurements over the full orbit from observations on MJD 58671 and the bottom panel is for an observation made 9 days later (MJD 58680).
\label{fig:3}}
\end{figure}

\section{Analysis and Results}
\subsection{Precursor emission at low frequencies}
%Millisecond pulsars (MSPs) are known for their remarkable pulse profile evolution \citep[e.g.,][]{Dai2015,Bhat2018}. 
%Revised RB
Some millisecond pulsars (MSPs) are known for their remarkable pulse profile evolution \citep[e.g.,][]{Dai2015,Bhat2018}. 
Unlike other bright southern MSPs (such as PSRs J0437$-$4715 and J2145$-$0750), PSR {\PSR} has a narrow pulse profile and shows very little profile evolution from 100\,MHz to 4\,GHz, as described earlier. As seen from Figure \ref{fig:1}, overall, the main profile shows very minimal evolution above 500\,MHz. At frequencies below 500\,MHz, there is clear evidence of the presence of additional precursor components, which were first seen 
at frequencies below 300 MHz with the MWA
\citep{Kaur2019}. Our observations in the uGMRT Band 3 have now unambiguously confirmed the presence of this evolving precursor emission, as evident in Figure \ref{fig:1}.

Based on our previous results and a non-detection of precursor emission at Parkes frequencies, we had earlier suggested that the precursor emission may have a steeper spectrum (spectral index $\alpha$\textless\,${-}3.7$, where $\alpha$ is defined as flux density $S \propto \nu^{\alpha}$, and $\nu$ is  frequency) compared to that of the main pulse \citep{Kaur2019}. Even though our uGMRT data are not flux calibrated, assuming the nominal system parameters, i.e. the system temperature, $\rm T_{\rm sys}$ = 106\,K, and a gain $\rm G$ = 0.32\,K/Jy \citep{Gupta2017}, we estimate $\sim$0.1\,mJy for the rms noise in uGMRT Band 3, accounting for the number of antennas (26) used in that observation. The precursor amplitude seen in uGMRT Band 3 is $\sim5\%$ of the main pulse (see Figure \ref{fig:style_plots}). This would translate to a precursor peak flux density of $\sim$3\,mJy, which is consistent with \textless\,12\,mJy that we would expect based on the extrapolation from the MWA measurements (and assuming $\alpha < -3.7$; \citealp{Kaur2019}).

\subsection{DM Measurements}
%As evident from Figure \ref{fig:style_plots}, there is no significant profile evolution within any of the four bands as well as throughout the entire observed band from $\sim$100\,MHz to 4\,GHz (see also Figure \ref{fig:1}). 

As described in \S~2 and Table~1, observations were made over large bandwidths for all three telescopes (see also Figure \ref{fig:1}). 
Such large bandwidth observations would be sub-banded into smaller frequency chunks, 
%and TOAs would be generated from each subband in order to properly account for the profile evolution \citep[e.g.,][]{Pennucci2019}. 
%Revised RB
and TOAs would be generated from each subband in order to properly account for the profile evolution, as typically done for 
PTA analysis until recently.  
For PSR {\PSR}, even though there is some measurable profile variation from the MWA to UWL band due to precursor emission, the degree of variation of the main pulse component is very small. In fact, the measured pulse width at 10\% of the peak 
%($\rm W_{10}$) 
changes from the MWA to UWL band by about $\sim$10\% ($150\,\mu$s to $130\,\mu$s). 
%The degree of profile evolution within any single frequency band (as shown in Figure \ref{fig:style_plots}) is also very small ($\sim2$\%). 
%RB revised in response to the referee comment 
The degree of profile evolution within any single frequency band (as shown in Figure \ref{fig:style_plots}) is generally small; in terms
of measured pulse width (e.g., quantified using $W_{50}$, at 50\% of 
the peak), we measure a change of $\sim2$\% across the 
two uGMRT bands, $\sim$5\% across the MWA band, whereas it is $\sim$10\% across the large frequency
range of Parkes UWL (0.7 to 4 GHz).
%Given this, we have chosen a single template for each observing band; specifically, four different templates were used for the MWA, uGMRT Band 3, uGMRT Band 4, and Parkes-UWL, to incorporate the small degree of observed profile evolution.
%RB revised in response to the referee comment #1
In light of this, we adopted two different methods for our timing analysis and DM measurements: the first method involved the use of a single template for each observing band, which is akin to ``narrow-band" timing, as it does not account for any in-band pulse profile evolution (\S~4.2.1). The second method is essentially wide-band timing based on the pulse portraiture that was originally developed by \cite{Pennucci2014} and further discussed in \cite{Pennucci2019}. 

%Not sure if this should be here or further down? 
%As a further cross check, we used the {\tt pat} routine from {\tt PSRCHIVE} to examine the difference between the measured and template profiles, and to assess the `goodness' of our analytic templates. No prominent profile-like features were seen in the residuals of any of the observations (after subtracting the best-fit template), which further justifies our approach.

\subsubsection{Narrow-band timing using analytic templates}
For this analysis we used different templates for the MWA, uGMRT Band 3, uGMRT Band 4, and Parkes-UWL, in an effort to incorporate the small degree of observed profile evolution. For the UWL data, we chose the Parkes 20\,cm (1.4 GHz) template that we obtained from the PPTA project \citep{Kerr2020}. UWL data were integrated in frequency to 26 channels and then cross-correlated with the template to obtain the {\toa}s. For the uGMRT data (i.e., Band 3 and Band 4), we chose one of the brightest scans to generate noise free analytic templates by modeling a sum of von Mises functions using the {\tt paas} utility of {\tt PSRCHIVE}. The uGMRT data were then integrated in frequency to 32 channels and cross-correlated with respective analytic templates to obtain the {\toa}s. For the MWA data, we used an analytic template from one of our bright observations recorded in 2017 (also published in \citealp{Kaur2019}) 
%to generate an analytic template. 
The analytic template is then cross-correlated with the observed profiles to obtain the {\toa}s using the {\tt PSRCHIVE} package.

The {\toa}s were then analysed using the pulsar timing package {\tt TEMPO2} \citep{Hobbs2006} to determine the DM. 
We adopted the latest available timing solution of the pulsar 
(from the PPTA project; \citealp{Kerr2020}) and fit for only 
%the DM in this analysis. 
%Revised RB
the DM in this analysis.\footnote{The PPTA solution is from timing observations up till 2018, however timing ephemerides extrapolare well for this pulsar as our analysis is primarily on frequency-dependent effects.}  
As described in \cite{Kaur2019}, for most of our observations with the MWA, we typically achieve a timing precision of the order of $\sim1\,\mu$s. Timing precision of the order of $\sim$2\,$\mu$s was achieved with uGMRT Band 3; however, a better timing precision of $\sim$0.5$\,\mu$s was obtained with Band 4. For our Parkes analysis, timing precision ranged from $0.9\,\mu$s (at the low-frequency end of the UWL band) to $3.5\,\mu$s (at the high-frequency end of the UWL band), as expected owing to the typically steep spectrum of radio pulsars (cf. Figure 1). This timing precision is reasonable considering the use of multi-frequency arrival times to determine the DM.\\ 

\subsubsection{Wideband timing using the pulse portraiture}
This analysis made use of the {\tt PulsePortraiture} package, details of which are described in \citet{Pennucci2019}. We obtained a two-dimensional (2D) template from the PPTA project, which was generated by combining 10 very high-quality observations when the pulsar was exceedingly bright and RFI was minimal. 
%old
%The UWL band was sub-divided into three near-octave sub-bands, in a way similar to that of narrow-band timing analysis as described in \S~4.2.1, using {\tt pazi} package of {\tt PSRCHIVE}. 
%modified
For the DM analysis, UWL data were sub-divided into three near-octave sub-bands, using the {\tt pazi} routine within {\tt PSRCHIVE} package. DMs and TOAs were calculated for each of these three sub-bands.

For uGMRT Band 3 and Band 4 data, the 2D templates were generated from our own observations. For the MWA, we co-added three of our high-quality VCS observations to generate the 2D template. This was then used to measure the DM and TOAs for each of our MWA observations. The DM measurements from our analysis are summarised in Table \ref{tab:MWA_observations}.

The achievable DM precision  strongly depends on the observing frequency band. The low frequencies of the MWA enable the most precise DM measurements, typically of the order of $\sim$(2-6)$\times10^{-6}$\,{\DMunit}. These are comparable to our previously published results \citep{Kaur2019}, and are still amongst the highest precision DM measurements to date. However, at uGMRT Band 3, DM precision achieved is $\sim 10^{-5}$\,{\DMunit}, an order of magnitude lower than that achieved using the MWA. Owing to much lower levels of RFI in Band 4, and due to a fortuitous scintillation brightening, we were able to achieve a comparable precision of the order of $\sim 2\times10^{-5}$\,{\DMunit}. In the Parkes UWL band, the achieved DM precision varies from $10^{-5}$ to $10^{-3}$\,{\DMunit} across the low to high segments of the band.

\subsection{Orbital DM variations}
The high DM precision achievable with the MWA offers the prospects of investigating any DM variations as a function of the orbital phase. Figure \ref{fig:3} shows DM measurements across the full 3.5-hour orbital period of the pulsar obtained from our  MWA observations (cf. Table \ref{tab:MWA_observations}). The top panel shows the DM measurements over the full orbit from observations on MJD = 58671 and the bottom panel 
shows similar measurements made nine days later (MJD = 58680).
No significant DM variation is seen
as a function of orbital phase, except for a noticeable excess near orbital phase $\sim$ 0.4 - 0.5, where a DM change of the order of $\rm (1.4\pm0.6)\times10^{-5}$\,{\DMunit} is measured.

\begin{figure}
\gridline{\fig{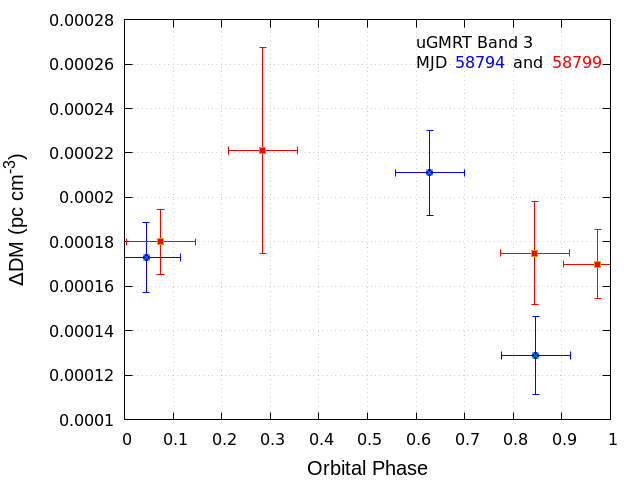}{0.45\textwidth}{}}
\vspace*{-0.5cm}
\caption{DM vs orbital phase from uGMRT (Band 3) observations. DM measurements are from two different epochs separated by five days (MJDs shown in the panel). The DM range in this panel is nearly 3 times larger than those in Figure 3 for MWA measurements. 
\label{fig:4}}
\end{figure}

Figure \ref{fig:4} shows similar measurements for the uGMRT data, which are from observations at two different epochs, separated by five days (MJD = 58794 and 58799). Even though the DM measurements with the uGMRT data are an order of magnitude less precise compared to the MWA measurements, and do not sample the exact same orbital phase ranges, there is a marginal trend for a slightly higher DM around $\sim$\,0.3 $P_b$ to 0.6 $P_b$ in the orbital phase. The uGMRT measurements also show larger DM variations on the order of $\rm \sim 10^{-4}$\,{\DMunit}. However, unlike the case with MWA observations, these observations were not made on the same day, and therefore these orbital DM variations can be difficult to disentangle from any temporal DM variations given the time span of our observations. On the other hand, the MWA observations were taken on the same day, and therefore give us confidence that the variations are loosely correlated with the binary orbital phase, and are not due to other (temporal) DM variations.

%\caption{DM vs orbital phase from uGMRT (Band 3) observations. DM measurements are from two different epochs separated by five days. Observations were planned in such a way that we covered different parts of the orbit on two different days. The y-range is 3 times larger than the full-orbit measurements from the MWA. This is possibly a result of combination of orbital and temporal DM variations.

\subsection{Frequency-dependent DMs} \label{s:freqdms}

%\begin{figure}
%\includegraphics[width=0.45\textwidth, height=0.25\textheight]{Fig5a_new.png}
%\includegraphics[width=0.45\textwidth, height=0.25\textheight]{Fig5b_new.png}
%\includegraphics[width=0.45\textwidth, height=0.25\textheight]{Fig5c_new.png}

%\includegraphics[width=0.45\textwidth, height=0.25\textheight]{07-09nov_PP.png}
%\includegraphics[width=0.45\textwidth, height=0.25\textheight]{12-13nov_PP.png}
%\includegraphics[width=0.45\textwidth, height=0.25\textheight]{29-30nov_PP.png}

%\caption{DM measurements of PSR {\PSR} at multiple frequency bands spanning the frequency range from 80\,MHz to 4.0\,GHz. Observations were made within 24-48 hour observing windows. The dashed line represents an empirical fit to a power law, and the solid gray line represents the reference DM value, $DM _{0}$, from the fit.
%The inset plot represents the frequency range $\sim$100 to $\sim$500\,MHz, to highlight the higher DM precision (of the order of $\rm \sim 10^{-6}$ to $10^{-5}$\,{\DMunit}) obtained using the MWA and uGMRT. Top and middle panels show measurements from the MWA, uGMRT Band 3 and the Parkes UWL receiver, with best-fit scaling indices (\S~\ref{s:freqdms}) $x=3.2\pm0.6$ and $x=2.9\pm0.4$, respectively. The bottom panel shows measurements from the MWA, uGMRT Band 4, and Parkes UWL, with $x=4.1\pm0.4$.}
%\label{fig:5}
%\end{figure}

\begin{figure*}
\gridline{\fig{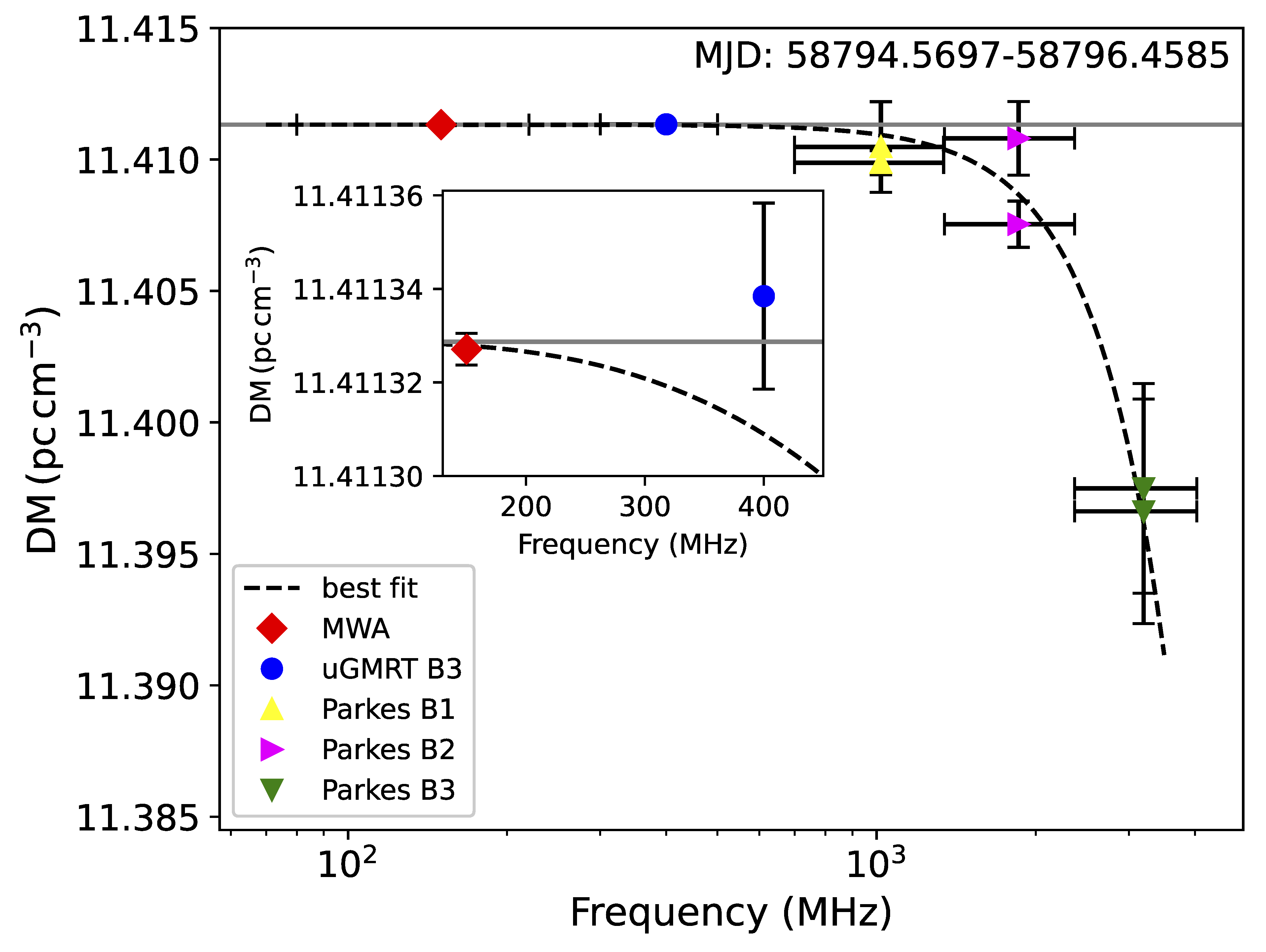}{0.42\textwidth}{}
          \fig{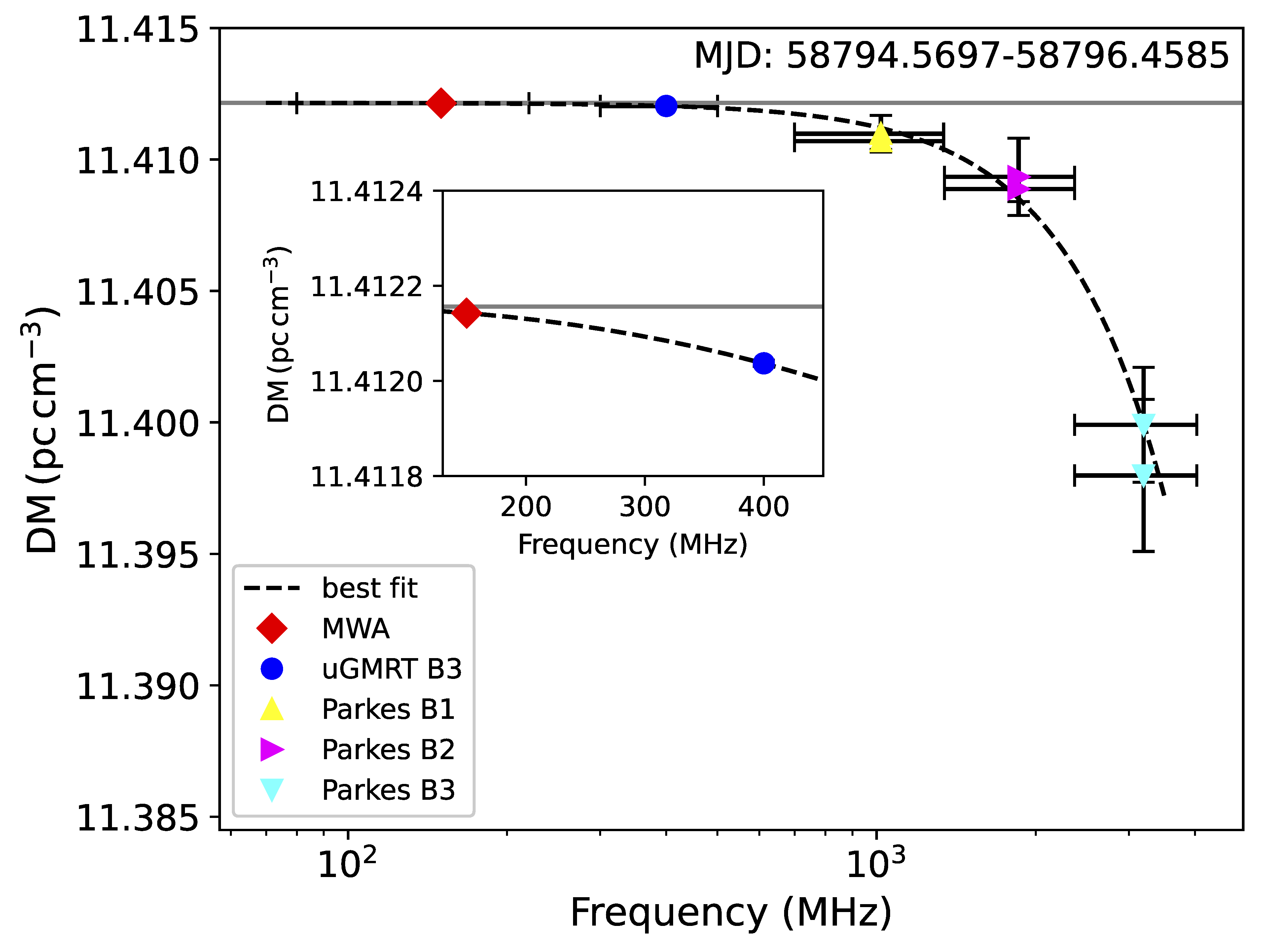}{0.42\textwidth}{} 
          }
\gridline{\fig{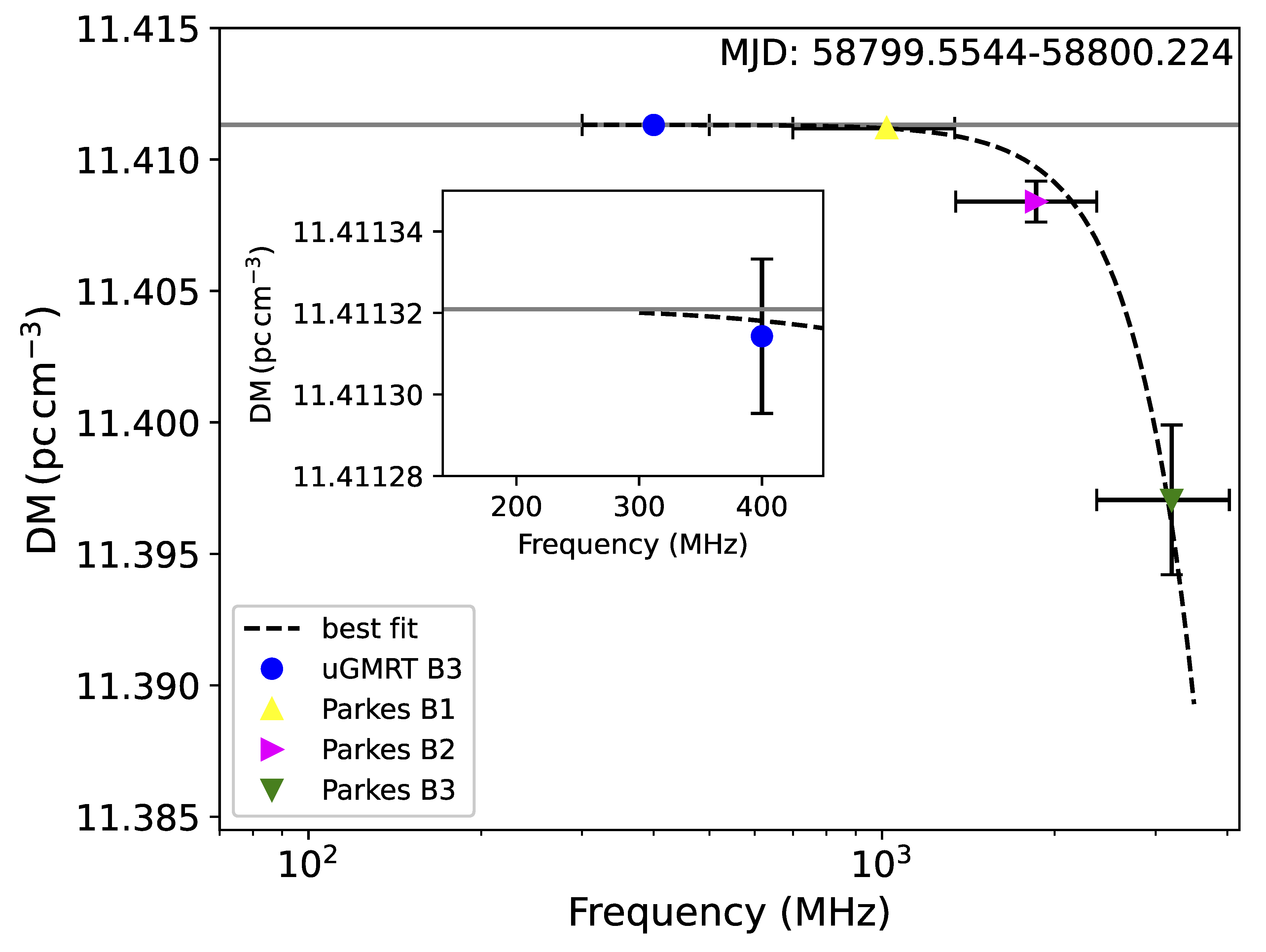}{0.42\textwidth}{}
          \fig{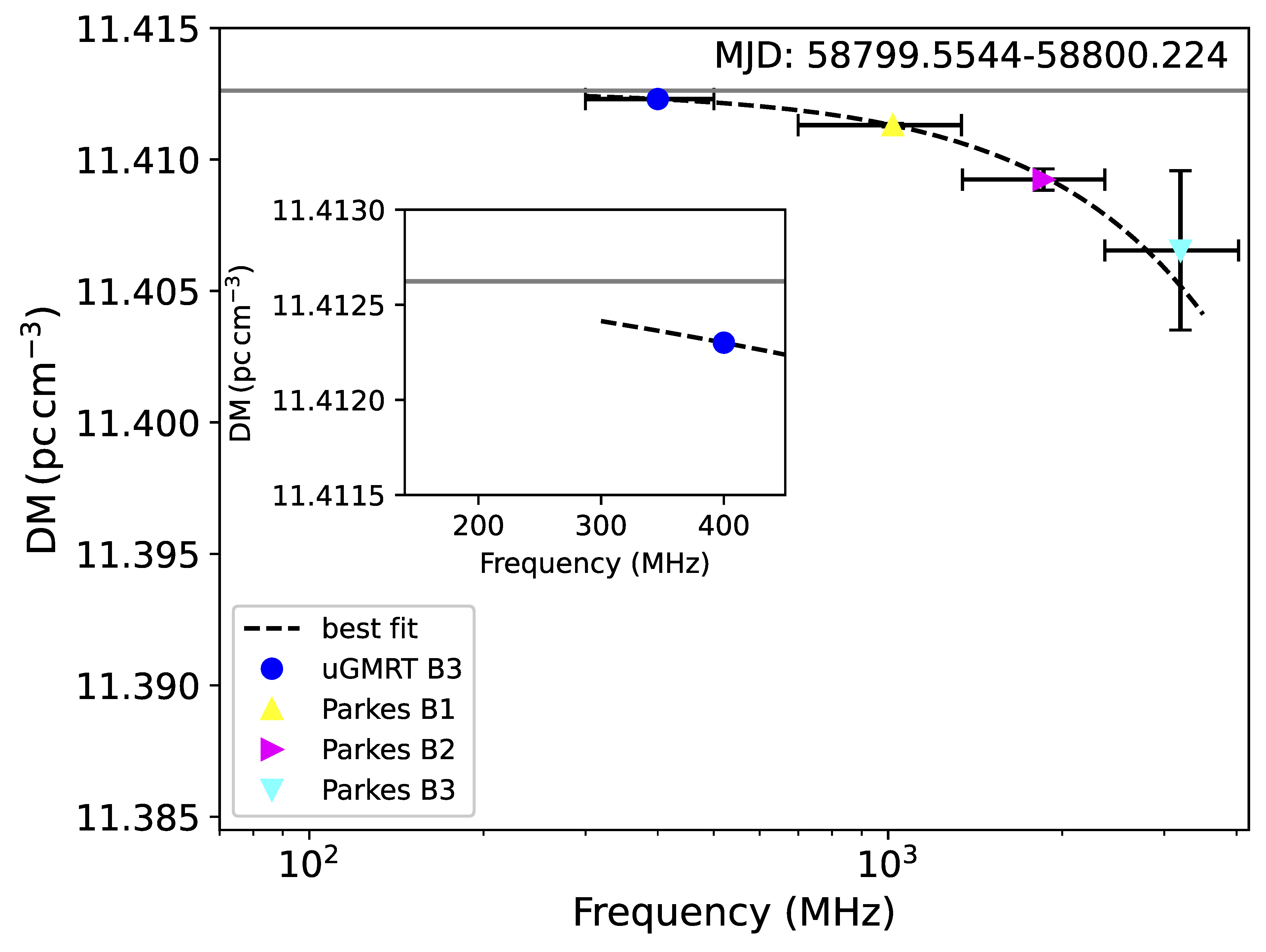}{0.42\textwidth}{} 
          }          
\gridline{\fig{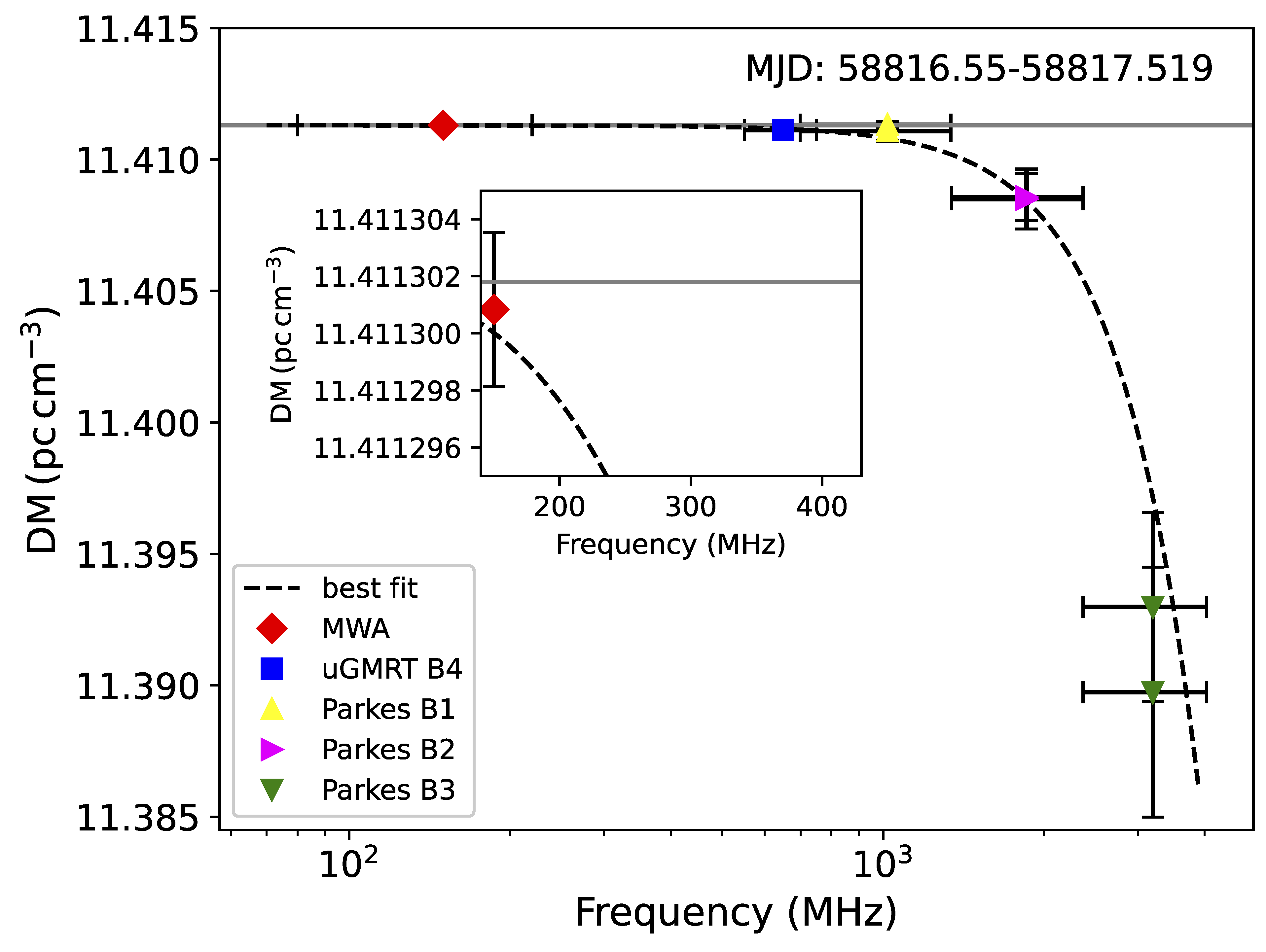}{0.42\textwidth}{}
          \fig{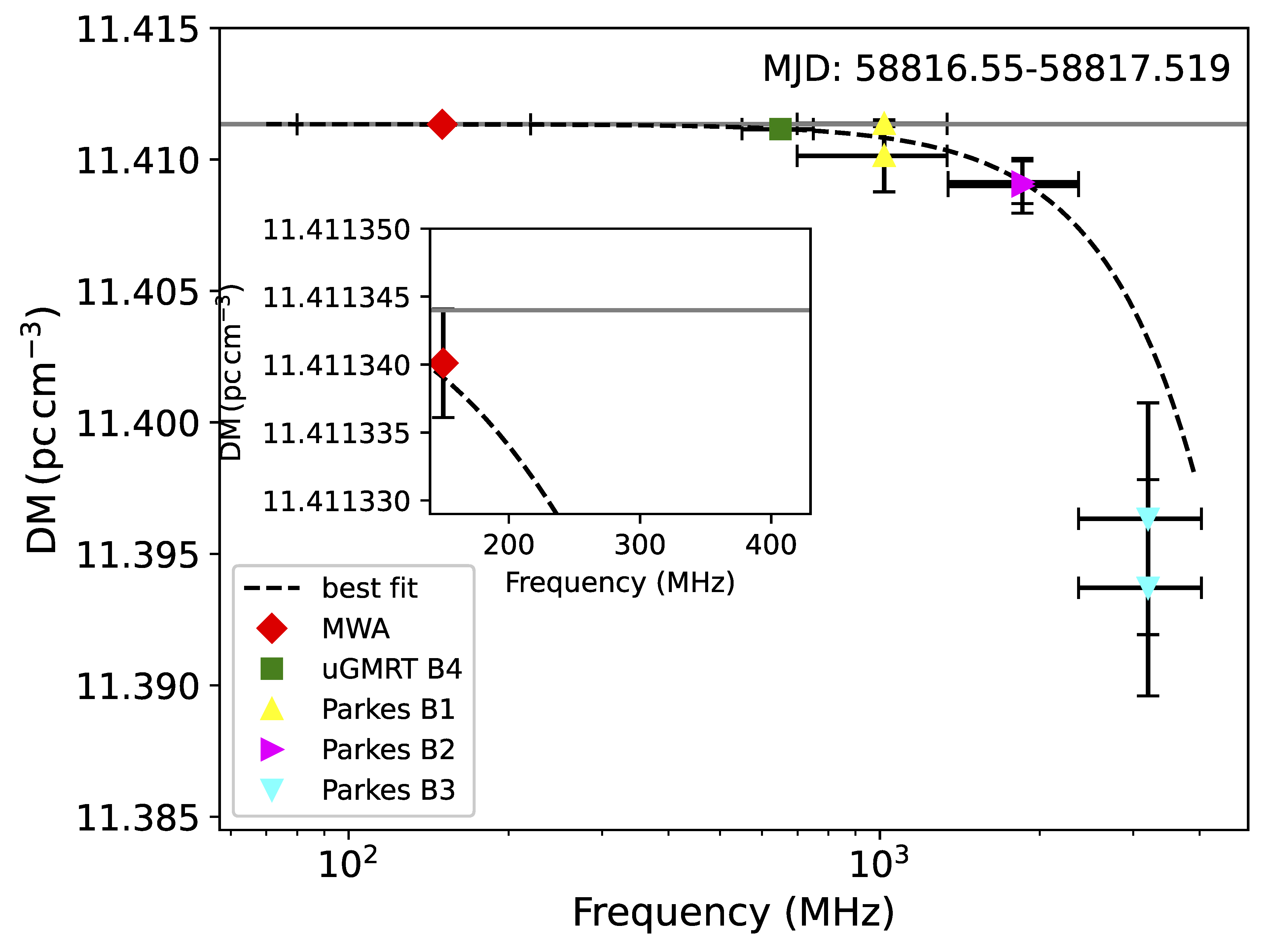}{0.42\textwidth}{} 
          }
\caption{DM measurements of PSR {\PSR} at multiple frequency bands spanning the frequency range from 80\,MHz to 4.0\,GHz. Observations were made within 24-48 hour observing windows. The dashed line represents an empirical fit to a power law, and the solid gray line represents the reference DM value, $DM_{0}$, from the fit. The inset plot represents the frequency range $\sim$100 to $\sim$500\,MHz, to highlight the higher DM precision (of the order of $\rm \sim 10^{-6}$ to $10^{-5}$\,{\DMunit}) obtained using the MWA and uGMRT. 
Left: DM values measured from narrow-band timing; Right: DM measurements from wide-band timing analysis.
Top and middle panels show measurements from the MWA, uGMRT Band 3 and the Parkes UWL receiver,
%with best-fit scaling indices (\S~\ref{s:freqdms}) $x=3.2\pm0.6$ and $x=2.9\pm0.4$, respectively. 
The bottom panel shows measurements from the MWA, uGMRT Band 4, and Parkes UWL. The measured scaling indices and $\rm DM_{0}$ values are presented in Table \ref{tab:fit-summary}. \label{fig:5}}
%with $x=4.1\pm0.4$.} \label{fig:5}
\end{figure*}

Figure \ref{fig:5} presents DM variations over the large frequency range from 80\,MHz to 4\,GHz. 
The three different panels shown here are for different data sets (i.e. from different epochs). 
For example, the MWA, uGMRT Band 3 and UWL were available for observations made 
on MJD 58794-58796; the uGMRT Band 3 and UWL were used for observations on MJD 58799-58800; and the MWA, uGMRT Band 4 and UWL were used for observations on MJD 58816-58817.

Our data sets clearly indicate that  DM changes measurably across the large frequency range. This frequency dependent behavior is seen consistently at all three epochs. We attempted to model this empirically using a power-law of the functional form 
$\delta DM \propto \nu^{x}$, where $\delta DM = DM_{\rm meas} - DM _{0} $; 
$ DM _{\rm meas} $ and $DM _{0}$ denote the measured and reference DMs, respectively,  
and $x$ is the scaling index (i.e. the change in DM  relative to a reference value $ DM_{0} $, which denotes the asymptotic value near the low end of the frequency range of measurements). 
We measured scaling indices of $3.2\pm0.6$, $4.1\pm0.4$, $2.9\pm0.4$ from narrow-band timing and $2.2\pm0.1$, $1.5\pm0.1$, $2.4\pm0.5$ from wide-band timing at three epochs, i.e. MJD = 58794, 58799 and 58816, respectively. We note that the fits are dominated by the low-frequency DM measurements which have uncertainties orders of magnitude smaller than the high-frequency DM measurements. An average value of $11.411318(6)$\,{\DMunit} and 11.412391(5)\,{\DMunit} was obtained for $ DM_{0}$ from two different timing analysis, compared to the catalog value of $11.41151(2)$\,{\DMunit} (from \citealp{Kaur2019}).

Evidently, the DM measured at low frequencies ($ \lesssim \,1\,$GHz) is significantly different 
to (in this case, higher than) those measured at higher frequencies ($\gtrsim$\,1\,GHz). However, 
it may be possible to measure a predictive scaling, which can be applied to make meaningful DM 
corrections in PTA observations.  Whether such a scaling is pulsar dependent, or epoch dependent, 
requires further investigation.

%%%%%%%%%%%%%%%%%%%%%%%%%%%%%%%%%%%%%%%%%%%%%%%%%%%%%%%%%%%%%%%%%%%%%%%%%%%%%%
% (revised text - RB)

\begin{deluxetable}{ccccc}
\tablenum{3}
\tablecaption{Summary of DM fit parameters. \label{tab:fit-summary}}
\tablehead{\colhead{MJD} & \multicolumn{2}{c}{Scaling index} & \multicolumn{2}{c}{${\rm DM_{0}}$\,({${\rm pc\,cm^{-3}}$})} \\
{} & AT & PP & AT & PP}
\startdata
58794-58796 & $3.2\pm0.6$ & $2.2\pm0.1$ & 11.411329(5) & 11.41262(6) \\
58799-58800 & $4.1\pm0.4$ & $1.5\pm0.1$ & 11.41132(3) & 11.412156(6)\\
58816-58817 & $2.9\pm0.4$ & $2.4\pm0.5$ & 11.411302(8) & 11.41134(1)\\
\enddata
\tablecomments{AT and PP denote the measured scaling indices from narrow-band and wide-band timing, respectively. $\rm DM_{0}$ is the reference DM value.}
\end{deluxetable}

\section{Discussion}
Our analysis suggests a clear, and quite compelling, evidence that the DM toward the millisecond pulsar {\PSR}, as measured in observations, depends on the observing frequency range spanned. 
%It is seen consistently at three independent observing epochs spanning about a month.
%RB revised
It is seen consistently using two different timing analyses, and across three independent observing epochs spanning about a month.
The observations were made contemporaneously using three different telescopes, covering a large frequency range from 80\,MHz to 4\,GHz. Specifically, in the case of narrow-band timing, we measure a $\rm \delta DM$ relative to a reference DM, $DM_{0} $,
%of 11.41151\,{\DMunit}, 
the magnitude of which depends on the frequency range sampled;  specifically, we measure $\rm \delta DM \sim (1.2 \pm 0.2) \times 10^{-4}$\,{\DMunit} across 0.1-0.5\,GHz, and $\rm \delta DM \sim (1.5 \pm 0.3) \times 10^{-2}$\,{\DMunit} across 0.1-3\,GHz, whereas the corresponding values from wide-band timing are $\sim (1.1 \pm 0.1) \times 10^{-4}$\,{\DMunit} and $\sim (1.4 \pm 0.3) \times 10^{-2}$\,{\DMunit}, respectively. This change in measured DM scales with the observing frequency ($\nu$) as $\rm \delta DM \sim \nu^{3.4 \pm 0.3}$ for narrow-band timing, and $\rm \delta DM \sim \nu^{2.2 \pm 0.1}$ for wideband timing, where the quoted scaling index is the mean value of the indices estimated for three different observations.

%either here or in Sec 4.4
The scaling indices derived from wideband timing measurements are shallower than those using analytic templates. They are also generally consistent within $\sim$1-2\,$\sigma$ except for those on MJD 58799-58800, when there were no MWA observations, suggesting that low-frequency measurements are crucial in constraining the scaling indices.

%revised following MB comment
%{\bf 
This presents the first reported evidence that the measured DM varies with the observing frequency, and that the change in DM scales with frequency but in a quantifiable way; this is inferred from contemporaneous observations over a large frequency range from 80 MHz to 4 GHz.
%}
Since PSR {\PSR} is amongst the most promising targets for current and future PTAs, 
this finding has important implications. 
We consider various plausible sources of DM variations applicable to such high-precision timing, such as  temporal variations in DM, or those caused by propagation effects such as scattering, as well those that may arise from intrinsic or environmental properties, and argue that none of them is relevant or can account for the observed changes in DM with frequency. 

Temporal variations in DM caused by  large space velocities of pulsars  ($\sim$80\,${\rm km\,s^{-1}}$ for PSR~{\PSR}, based on recent proper motion measurement and assuming a distance of 0.96 kpc; \citealp{Reardon2021}) is a dominant source of noise in PTA measurements. While there are no published records of longer-term DM variations for this pulsar, our MWA data suggest $\rm \delta DM \sim 10^{-4}$\,{\DMunit}, on timescales of $\sim$1-2 yr \citep{Kaur2019}.
The fact that our observations were made contemporaneously, through careful   coordination of the three telescopes within $\sim$24-48\,hour time windows, naturally mitigates this effect.   

Since PSR {\PSR} is in a 3.5-hr binary orbit with a black-widow type companion, another possibility to consider 
is variation of DM with orbital phase.
Even though the pulsar does not show any evidence of eclipse \citep{Keith2011}, in light of the orbital modulations seen in high-energy gamma-ray observations and its interpretation in terms of intra-binary shock emission \citep{An2018}, it was imperative to consider (and investigate) this effect. However, as discussed in \S~4.2.1, the measured $\delta$DM from our specially designed MWA observations is rather small, i.e. $\delta$DM of $\sim \rm 1.4\times10^{-5}$\,{\DMunit}, which is $\sim$1-2 orders of magnitude smaller than measured DM changes across the frequency range. This therefore readily precludes any orbital variations. Even so, this small change in DM will result in a timing delay of $\sim3\,\mu$s at the MWA's observing band (120-220\,MHz), which will translate to $\sim110$\,ns in the Parkes UWL band (700-4000\,MHz). 

Another important effect to consider is multi-path propagation that can give rise to a scatter broadening of the observed pulse shape, which varies strongly with frequency.
For PSR {\PSR}, our observations do not reveal any visible scatter broadening even at the lowest frequencies of our observations (at 80\,MHz). Closer examination of the data presented in Fig. \ref{fig:style_plots} suggests a scintillation bandwidth $\sim$\,5\,MHz at 400\,MHz (i.e., uGMRT Band 3), which translates to a pulse broadening time ($\tau _d$) of $\sim$\,30\,ns at this frequency, and $\sim$\,20\,$\mu$s at the MWA's 80\,MHz (assuming a scaling of $\tau _d \propto \nu^{-3.9}$ as per \citealp{Bhat2004}). 
Notably, the expected pulse broadening time is $\lesssim$1\,$\mu$s at frequencies $\gtrsim$ 200 MHz, which is significantly smaller than $\sim$5 $\mu$s dispersive time delay expected for a measured  $\delta DM \sim 1.2 \times 10^{-4}$\,{\DMunit} across 0.1 - 0.5 GHz, and a much larger $\sim$2.5\,ms for $\delta DM \sim 1.4 \times 10^{-2}$\,{\DMunit} across 0.1-3\,GHz.
Even at the lowest frequency (80 MHz), the estimate of $\tau _d$ is still smaller than the dispersive delay $\sim$\,75\,$\mu$s across the MWA band (80-220 MHz) due to the measured change in DM. Therefore, frequency-dependent DMs seen in our data are unlikely due to variable scatter broadening across the frequency range.

% RB revised - as much of the below text is no longer relevant! (in light of new analysis) 
%To investigate possible DM measurement errors due to profile evolution, we tested our data using the available software packages that are designed to take into account the profile evolution with frequency. In particular, we used the {\sc PulsePortraiture} \citep{Pennucci2019} software package.  The DM values estimated using {\sc PulsePortrature} do not agree with those estimated using analytic templates as described in Section 4.2.  The {\sc PulsePortrature} DM estimates also exhibit more noise (around 4-5\,$\sigma$ of scatter), whereas the DM estimates derived using analytic templates show much less scatter (around 1-2\,$\sigma$). Therefore, in this paper, we have adopted the DM estimates derived using analytic templates. We also tested our data with a single frequency template in order to check if the frequency-dependent DM trend still persists; initially, we used the 20\,cm (1.4 GHz) template, then the 50\,cm (732 MHz) and 10\,cm (3.1 GHz) templates, all of which are from the PPTA project. All of these analyses produced similar results, which further strengthens the claim that our analysis is not significantly biased by any un-modeled profile evolution (which appears to be almost negligible for this pulsar across the entire observing band). The frequency-dependent DM variations are significantly higher than the measured DM uncertainties.

%new text
The qualitative agreement of results from two distinct timing analyses, as summarised in \S~4.2.1 and 4.2.2, clearly suggests that our analysis is not significantly biased by any un-modeled pulse profile evolution. As noted in \S~4.2, the observed profile evolution across the large frequency range is comparatively small for this pulsar, with $\sim$10\% change in the measured pulse width (quantified in terms of $W_{50}$) across 
80\,MHz - 4\,GHz spanned by observations. 
An attempt to fit for FD parameters yielded a marginally significant ($\sim$1.5\,$\sigma$) value for FD1, with $\sim$10\% improvement in the reduced chi-squared. 
Furthermore, the measured changes in DM values (i.e. 
$\delta DM$ relative to $DM_0$) are significantly higher than the uncertainties in our estimated DM values, as evident in Fig.~5. The wide-band timing analysis method using the {\tt PulsePortraiture} is designed to model and account for any pulse profile evolution. Given this, the remarkable consistency seen in terms of an implied frequency dependence between the two timing methods, across all three data sets, precludes (or at least significantly diminishes) the possibility that profile evolution is significantly influencing our analysis and results.

%additional text following MB comment -- but do we need this? or can we tone it down? 
%{\bf 
%While the above analysis suggests that the profile evolution is unlikely to be the cause of the DM dependence in frequency that we have observed, it is still possible that even a small degree of profile evolution may give rise to an `offset' contribution to the measured DM that can also be frequency-dependent. Such an effect will most likely depend on the pulsar, and to certain extent, also on the instrumentation employed and even the method used for the DM estimation. 
%}

%RB - slightly revised text 
While the profile evolution is unlikely to be the cause of an observed frequency dependence in DM measurements, any residual unmodelled profile evolution may give rise to an `offset' contribution to the measured DM that may still be frequency-dependent. Such an effect will most likely depend on the pulsar, and to a certain extent, also on the instrumentation employed, and even the method used for the DM estimation. In any case, we expect this to be a rather subtle effect for this pulsar.

Another possibility is that there may also be very subtle aberration and retardation (AR) effects which can impact DM measurements.
%----------------------------
%% Original (submitted) text:
%However, for PSR {\PSR}, the radius of the light cylinder is $\sim 100\,$km.
%Therefore, this effect is negligible under reasonable assumptions of emission heights and radius-to-frequency mapping for MSPs, and would only become significant in more slowly rotating pulsars with much greater light cylinder radii \citep{Blaskiewicz1991,Dyks2015a}.
%----------------------------
%% SM: Replacement text
However, AR effects, which depend on the emission height and the radius-to-frequency mapping, are not expected to change over time, as these quantities are functions of the viewing geometry and the magnetic field configuration.
Therefore, AR effects cannot be responsible for the difference between our measurements of the DM (for instance), and historical DM measurements at the same frequency.
Thus, although AR effects could explain the observed $\delta$DM scaling at the reported epoch (at least qualitatively), we reject this explanation on the grounds that the observed effect (e.g., the measured scaling index) appears to be time-dependent.
%----------------------------

Finally, there are recent reports of this pulsar exhibiting astonishingly 
low levels of ``jitter noise," which has also been shown to contribute to the variability to DM estimates \citep[$\sim$\,4\,ns in an hour of observation;][]{Parthasarathy2021}; this result further strengthens the confidence in our DM estimates and an argument in support of a frequency dependence in DM.

%Revised RB [new text, in response to the referee comment]
\cite{Cordes2016} present a detailed account of frequency-dependent DMs including their expected magnitudes and scaling in frequency as a function of the observing frequency. According to their Equation (12), we may expect a predicted difference in DM between 150 MHz and 3.2 GHz of the order of $\sim  3\times10^{-6}$\,{\DMunit}, assuming a scattering screen placed half-way between the pulsar and us, and a pulsar distance of $\approx$1 kpc \citep{Reardon2021}, and scattering strength consistent with our measurement of the diffractive scintillation bandwidth (Fig 2; panel 2). We note that this is however much smaller than our measured difference in DM of $\sim$ 0.01\,{\DMunit}. Furthermore, the difference in DM increases with an increase in frequency, which is in contrast with the theoretical predictions. The analysis and results presented in this paper motivate further detailed studies of frequency-dependence in DM toward other PTA pulsars.

\section{Summary and Conclusions}
In this paper we have presented the results from contemporaneous observations of PSR {\PSR} made with the MWA, uGMRT and Parkes at three epochs, spanning about a month. Our observing and analysis strategies have been devised so that effects such as variations of DM with time, or orbital phase, can be either measured or minimised before frequency-dependence in DM can be investigated. 
This is the first time that this pulsar has been observed over such a wide-frequency range. Our measurements demonstrate that the MWA enables estimation of DMs with precision of the order of $10^{-6}$\,{\DMunit}, which is much better than that typically obtained from timing-array observations.

%revised following MB comment 
%{\bf 
We have reported the first convincing evidence that the DM as measured in wideband observations, depends on the frequency range spanned. This is demonstrated using broadband data obtained for PSR {\PSR}, and is seen consistently at three different observing epochs, and using two different timing techniques. 
%}
Effects of this kind, if confirmed for other PTA pulsars, will have important implications for timing-array experiments. 

Our results cannot be attributed to temporal or orbital variations, or other effects such as profile evolution with frequency, and suggest 
a {\em bona-fide}
frequency-dependence in measured DMs. 
The fact that DMs measured at low frequencies differ from those measured at high frequencies suggests that low-frequency observations cannot be straightforwardly used to apply DM corrections in PTA observations.
The larger fluctuations in DMs at higher frequencies might be attributed to averaging effects caused by the larger sizes of scatter broadened images at lower frequencies \citep{Cordes2016,Shannon&cordes2017}.

Our analysis shows that it is possible to empirically estimate the scaling index for frequency-dependent DM, 
in the same way that it can be done
for other frequency-dependent effects (e.g., scintillation or pulse broadening). 
Whether the scaling is constant for a given pulsar, or is epoch dependent, needs further investigation.
Regardless, for a given pulsar, if such a scaling can be verified, e.g., via suitably designed monitoring campaigns, low-frequency pulsar observations may still have an important role to play in applying effective DM corrections for pulsar timing arrays. 

\section{acknowledgments}
We thank the referee for their constructive comments that helped improve the quality of the results presented in this paper. This scientific work makes use of the Murchison Radio-astronomy Observatory, operated by CSIRO. We acknowledge the Wajarri Yamatji people as the traditional owners of the Observatory site. This work was supported by resources provided by the Pawsey Supercomputing Centre with funding from the Australian Government and the Government of Western Australia. The uGMRT is operated by the National Centre for Radio Astrophysics of the Tata Institute of Fundamental Research, India. The Parkes radio telescope (\emph{Murriyang}) is part of the Australia Telescope National Facility which is funded by the Australian Government for operation as a National Facility managed by CSIRO. DK acknowledges the support received from a Curtin International Postgraduate Research Scholarship award and a Curtin University Publications Grant. 
RMS acknowledges support through Australian Research Council Future Fellowship FT190100155.
The authors thank Matthew Bailes, Andrew Cameron, George Hobbs, Daniel Reardon, Marcin Sokolowski, Abhimanyu Susobhanan, Matthew Miles and Małgorzata Curyło for useful discussions.
%-----Need to add more----\\
%\vspace{5mm}
\facilities
The MWA, uGMRT, and Parkes radio telescope facilities.
\software 
This work made use of the following software packages: DSPSR \citep{van2011DSPSR}, PSRCHIVE \citep{Hotan2004, Van2012}, TEMPO2 \citep{Hobbs2012}, PulsePortraiture \citep{Pennucci2019}.

\bigskip

%\bibliography{sample63}{}
\bibliographystyle{aasjournal}

\end{document}